\documentclass[twocolumn,preprintnumbers,superscriptaddress,amsmath,amssymb]{revtex4}

\usepackage{graphicx}
\usepackage{dcolumn}
\usepackage{bm}
\usepackage{color}
\usepackage{epstopdf}
\usepackage{hyperref}

\begin{document}


\title{Stark spectroscopy of Rydberg atoms in an atom-ion hybrid trap} 
\author{Shinsuke Haze}
\affiliation{Institut f\"{u}r Quantenmaterie and Center for Integrated Quantum Science and Technology IQ$^{ST}$, Universit\"{a}t Ulm, 89069 Ulm, Germany}
\author{Joschka Wolf}
\affiliation{Institut f\"{u}r Quantenmaterie and Center for Integrated Quantum Science and Technology IQ$^{ST}$, Universit\"{a}t Ulm, 89069 Ulm, Germany}
\author{Markus Dei\ss}
\affiliation{Institut f\"{u}r Quantenmaterie and Center for Integrated Quantum Science and Technology IQ$^{ST}$, Universit\"{a}t Ulm, 89069 Ulm, Germany}
\author{Limei Wang}
\affiliation{Institut f\"{u}r Quantenmaterie and Center for Integrated Quantum Science and Technology IQ$^{ST}$, Universit\"{a}t Ulm, 89069 Ulm, Germany}
\author{Georg Raithel}
\affiliation{Department of Physics, University of Michigan, Ann Arbor, MI 48109, USA}
\author{Johannes  Hecker Denschlag}%
\affiliation{Institut f\"{u}r Quantenmaterie and Center for Integrated Quantum Science and Technology IQ$^{ST}$, Universit\"{a}t Ulm, 89069 Ulm, Germany}

%
%
%

\begin{abstract}

We report on Rydberg spectroscopy of ultracold atoms in an atom-ion hybrid trap for probing the electric fields in a mixture of atoms and ions. We obtain spectra which exhibit excitation gaps corresponding to 
avoided level crossings in the Stark map. From these measurements we can conclude that the ground state atoms experience electrical fields of up to 250~V/cm. There is, however, 
a difficulty in interpreting the results, because some data indicate that the electrical fields are produced by the ions while other data indicate that they
stem from the Paul trap. 
We discuss possible scenarios for explaining the measured data, provide first measurements to check these scenarios, and propose methods to finally solve this puzzle.
\end{abstract}
\maketitle

In recent years, a new research field has come up where cold ions and ultracold neutral atoms are experimentally brought together to study  their interactions  in  detail, for reviews see 
\cite{Haerter2014a, Willitsch2015, Tomza2017, Sias}.
In a number of experiments the interactions and reactions between cold atoms and ions have been investigated, e.g.
\cite{Grier, Rellergert, Ratschbacher, Hall1,Sivarajah, Ravi,  Sikorsky, Deiglmayr, Krukow, Hauser, Joger, Haze}. 
There is a considerable interest for tuning interactions, e.g. for controlling chemical reactions.
By   exciting atoms  or ions to electronically excited states  chemical reactions have  been influenced, see e.g. \cite{Hall2, Ratschbacher, Mills2019, Li2019}.
A particularly interesting  way for interaction tuning is coupling the atomic ground state to Rydberg states, making use of their  extraordinary properties \cite{Secker1, Secker2,Wang}. 
Recently, interactions between Rydberg atoms and ions have been observed \cite{Engel, Ewald}.
We report here on related experiments where we carry out Rydberg spectroscopy on a  trapped ensemble of ultracold  neutral Rb atoms and Rb$^{+}$ ions in the vicinity of the $27P$-state.  
The idea of this experiment is to probe atoms as they collide with ions. As atom and ion approach each other the electrical field of the ion  shifts the atomic Rydberg states into resonance 
with the spectroscopy laser light due to the Stark effect.
As a consequence, Rydberg states can be excited which are subsequently detected.  
 We observe excitation gaps in the Rydberg spectrum which correspond to 
 avoided crossings of the Rb Stark map.  These measurements show that the Rb atoms in the experiment experience 
electric field strengths of up to 250~V/cm. Concerning the origin of these fields, however, we obtain contradictory 
results. Several different measurements indicate that the electrical  fields stem from trapped ions, but comparison of our measured Rydberg spectra with Stark map calculations seem to indicate that the electrical fields might originate from the Paul trap.

\begin{figure}
	\includegraphics[width=8cm]{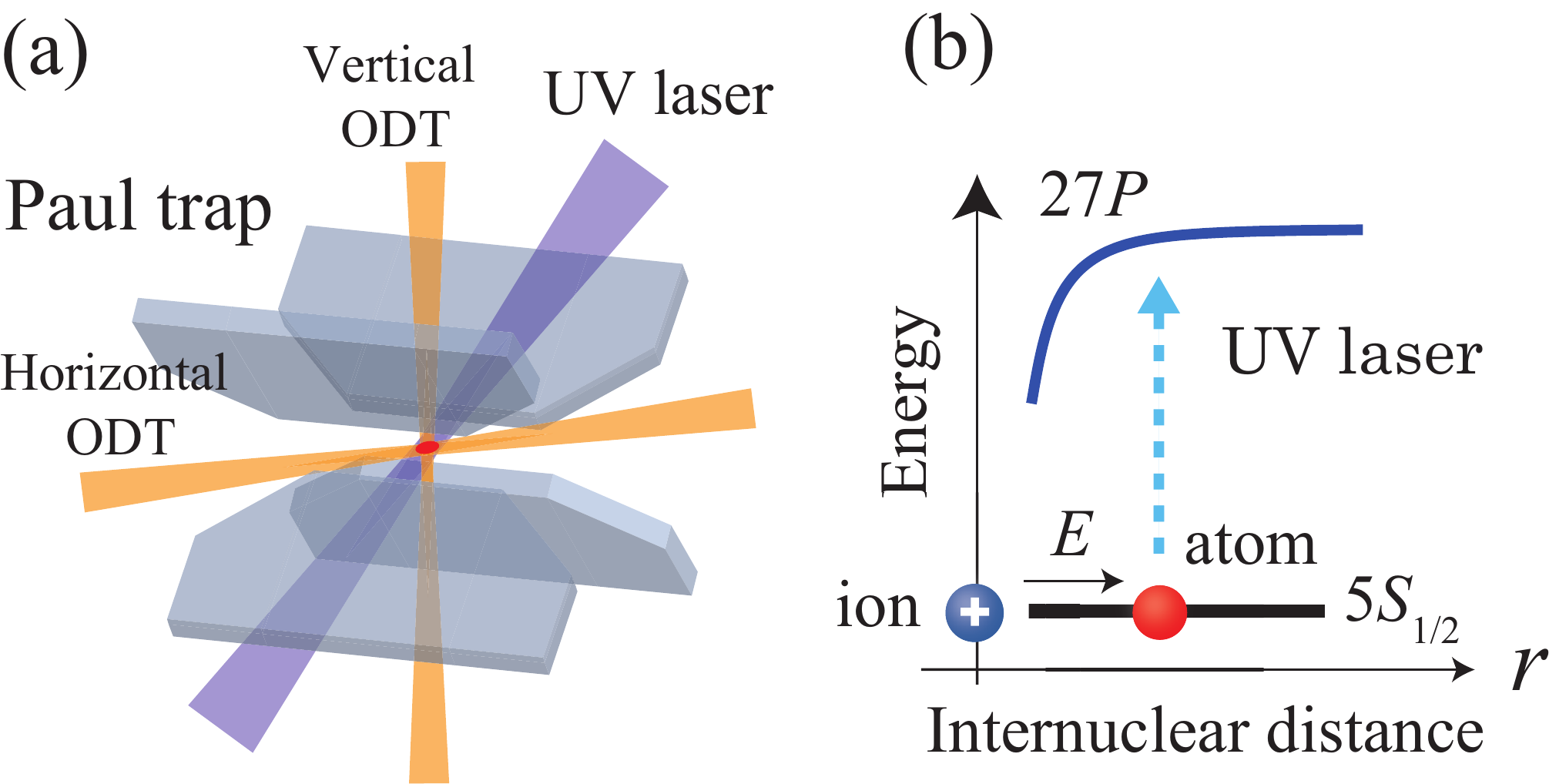}
	\caption{\label{Setup} (a) Rydberg spectroscopy setup in the hybrid atom-ion apparatus. 
Shown are the RF electrodes of the Paul trap, but not the  endcap electrodes. The optical dipole trap laser beams are orange. The UV laser beam is purple. 
(b) Rydberg excitation of a $5S_{1/2}$ ground state atom towards the $27P$ state in the electrical field of an ion. The Stark effect shifts the $27P$ level into resonance at a particular distance.
	}
\end{figure}

Our experiments are carried out in a hybrid setup of an optical dipole trap (ODT) and a linear Paul trap,  allowing for simultaneous trapping of atoms and ions, see Fig.~\ref{Setup}(a).  A detailed description of the apparatus can be found, e.g., in \cite{Schmid}.
 We start our experiments by initially preparing an ultracold gas of $^{87}$Rb atoms in the crossed-beam ODT at $1064\:\text{nm}$. These atoms are spin-polarized in the hyperfine state $(F=1, m_F=-1)$ within the $5S_{1/2}$ electronic ground state. The sample has an atom number of about 3.4$\times$10$^{6}$ and a temperature of $860\:\text{nK}$. We work with ODT trapping frequencies of $2\pi\times24\:\text{Hz}$ in the axial direction and $2\pi\times180\:\text{Hz}$ in the radial direction. The cloud has a Gaussian shape with a size of $\sigma_{x,y,z}=(60,8,8)\:\mu\text{m}$ along the three directions of space and the peak atom density is 4.7$\times$10$^{13}$~cm$^{-3}$.

The linear Paul trap is centered on the ODT. It  is  driven by a RF voltage at a frequency of $2\pi\times 4.2\:\text{MHz}$. The resulting axial and radial trapping frequencies for a Rb$^{+}$ ion are $\omega_{\rm{a,r}}=2\pi\times(42, 207)\:\text{kHz}$, respectively. Residual DC electric fields are well compensated as described in \cite{Harter2}, i.e. excess micromotion is minimized.

For the Stark spectroscopy we address Rydberg states in the vicinity of the $27P$ level
 via  single-photon excitation, starting from the Rb ground state.
  A frequency-doubled dye laser provides the corresponding  ultraviolet (UV) light at a wavelength of about $298.5\:\text{nm}$. The laser light is delivered to the vacuum chamber by a multi-mode optical fiber and therefore  essentially unpolarized.  The power of the UV laser beam   is  about $1\:\text{mW}$ and  its waist at the location of the atom cloud is typically $150\:\mu\text{m}$. For the measurements we use a light pulse of  rectangular shape
  and $225\:\text{ms}$ duration.

Once an atom is excited to a Rydberg state it can be ionized e.g. via photoionization by the ODT laser.
It is then subsequently trapped by the Paul trap. At the end of an experimental run we infer
 the approximate number of accumulated ions.
  However, we do have the capability to detect single ions with unit probability. The ion number is extracted from the measured atom loss from the cloud after an experimental run. The loss is mainly
   due to elastic collisions with ions in which  atoms are kicked out of the shallow ODT, see \cite{Harter1,Harter2,Wolf}. We measure the atom loss  via absorption imaging.

 Stark spectroscopy  probes the local electrical field of each Rb atom in the atomic cloud. This electrical field can either stem from an ion in the vicinity or from the ion trap.
In Fig.\:\ref{Setup}(b) we sketch  the Stark shift of the  Rydberg level $27P$ as a function of the distance between atom and ion, considering only the electrical field of a single ion.
At a particular distance the UV laser can resonantly drive a transition to the Rydberg level.

\begin{figure}
	\includegraphics[width=8cm]{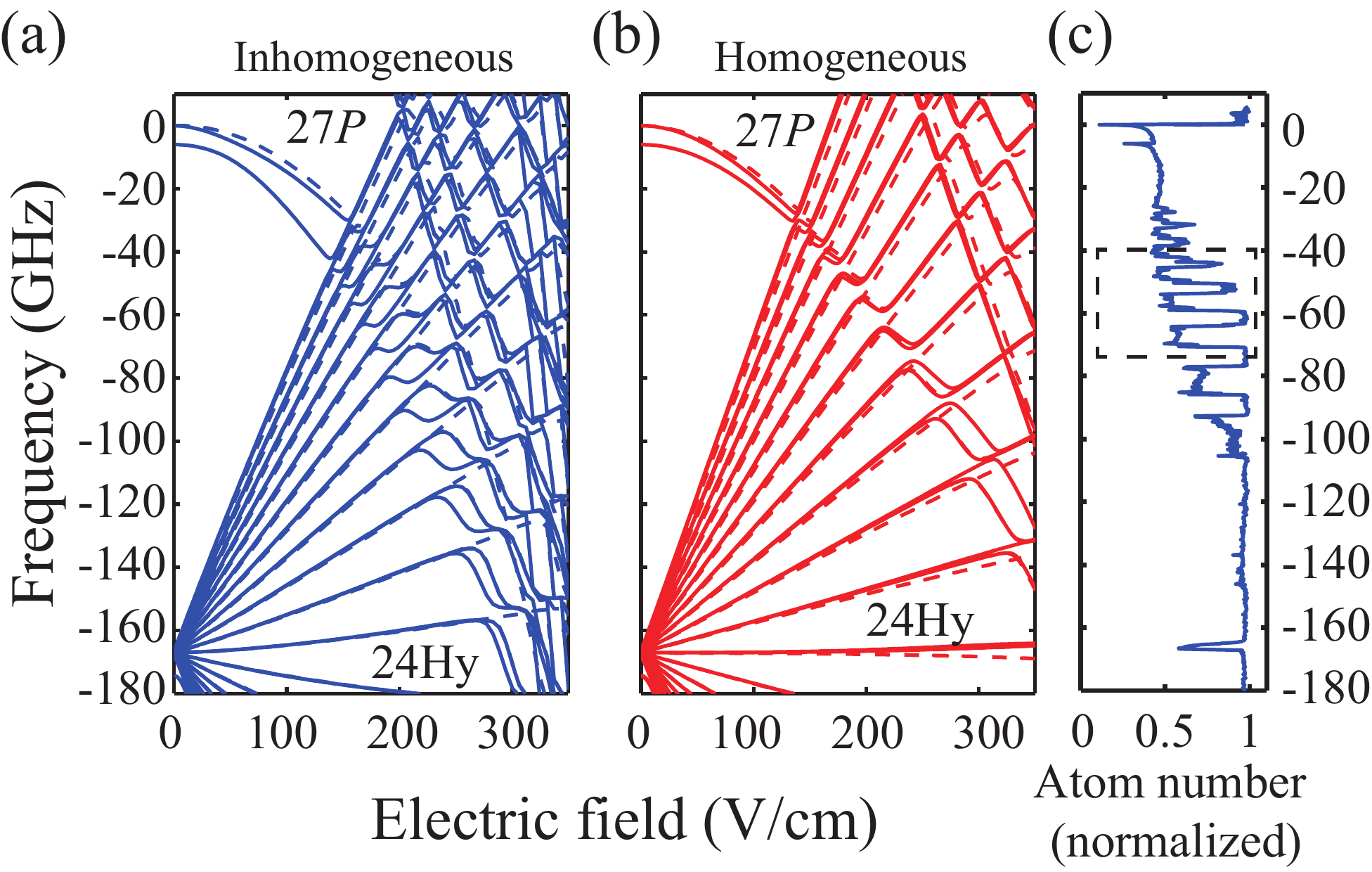}
	\caption{\label{n27_wide}  (a) and (b) Stark maps in the vicinity of the $27P$ state of $^{87}$Rb. 
The zero-field position of the $27P_{3/2}$ level serves as frequency reference. It corresponds to an offset
laser frequency of $1,004.48423\:\text{THz}$.
(a) Stark map for the inhomogeneous field of a point charge. (b) Stark map for a homogeneous field. Solid (dashed) lines indicate $|m_{j}|=1/2$ ($|m_{j}|=3/2$) states. Avoided level crossings occur between the $27P$ levels and high-$l$ states of the  hydrogenic manifold (24Hy).  (c)  Measured Rydberg excitation spectrum.
Each data point represents an average of 3 to 4 repetitions of the experiment. The black dashed box corresponds  to the zoom in Fig.$\:$\ref{zoom}.
	}
\end{figure}

Figure $\:$\ref{n27_wide}(a) shows the corresponding calculated Stark map. 
The $27P$ state is split into the fine structure components $27P_{1/2}$ and $27P_{3/2}$. 
Avoided crossings appear where the $27P$ levels cross the hydrogenic manifold 24Hy.
The shown electric field ranges from 0 to $350\:\text{Vcm}^{-1}$ corresponding to an atom-ion internuclear distance from $\infty$ to $203\:\text{nm}$.

The Stark map for the field of a point charge  is a distorted version of the standard Stark map for a homogeneous electrical field, see Fig.$\:$\ref{n27_wide}(b). 
Due to the large size of the Rydberg wavefunction and the huge polarizability the electrical field gradients have a non-negligible contribution to the Stark level shift. 
  
 Figure $\:$\ref{n27_wide}(c) is the measured Rydberg spectrum. 
 For this measurement the UV laser frequency was scanned with a stepsize of $100\:\text{MHz}$.
 For every frequency setting a freshly prepared Rb cloud was exposed to the 225ms laser pulse
 and subsequently the remaining atom number was measured. 
 The larger the loss, the larger the number of Rydberg atoms and ions that have been created. 
  The  ion production rate for a given laser frequency depends on several factors: the availability of resonant Rydberg levels, the transition matrix elements towards these levels, and the number of ground state atoms at the proper electric fields.  
 According to the electric dipole transition selection rules only the $27P$ state but not the high-$l$ states can be addressed when starting from the $5S_{1/2}$ ground state. However, close to  the avoided crossings $P$-state character is admixed to the respective high-$l$ state branch.
 We clearly observe two loss peaks from the $27P_{1/2}$ and $27P_{3/2}$ Rydberg resonances  at zero field. 
For increasing UV laser frequencies the loss signal quickly vanishes because of lack of $P$-states.
For decreasing  UV laser frequencies the loss signal decreases gradually. Although the two $P$-levels are still present, for resonant excitation increasing electrical fields are required and a decreasing number of atoms experience these stronger electrical fields.  
 
At the relative frequency of about -30 GHz, a strongly modulated, almost step like behavior of the measured signal occurs which coincides with the avoided crossings of the $P$-levels with the hydrogenic manifold. 
The modulation stops at a relative frequency of about -100 GHz which corresponds to an electrical field strength of around 250 V/cm.
Within the small atom cloud such a large field can only stem from ions. From the known parameters of our ion trap we calculate that the maximal electrical trap fields at the $1/e^2$-rim of the cloud are about 30 V/cm.

\begin{figure}
	\includegraphics[width=8cm]{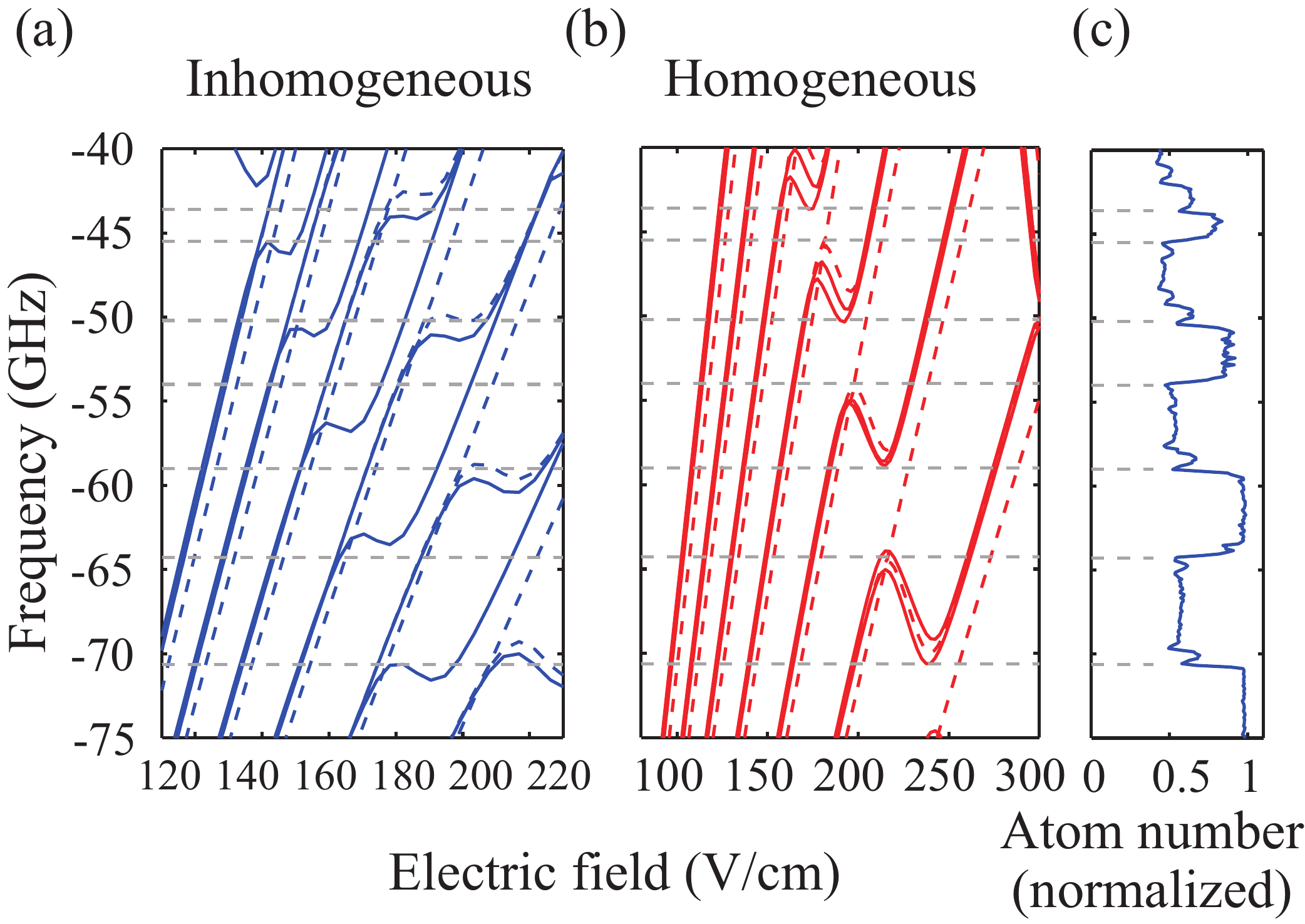}
	\caption{\label{zoom} Zooms into the plots of Fig.~\ref{n27_wide}, as indicated by the dashed box in Fig.~\ref{n27_wide}(c).
	 The dashed horizontal lines are  guides to the eye.  
	}
\end{figure}

 It turns out, however, that this conclusion is somewhat in conflict with further analysis of our spectroscopic data.
Figure $\:$\ref{zoom} is a zoom into the plots of Fig.~$\:$\ref{n27_wide}, corresponding to the dashed box in Fig.~$\:$\ref{n27_wide}(c).
When comparing the measured data with the calculated Stark maps it becomes clear that the Stark spectrum for homogeneous fields matches the data
much better than the Stark spectrum for the fields of a point charge. 
For example, there is  a very good correspondence of the  measured excitation gaps to the calculated  energy gaps at the avoided crossings for the Stark spectrum (b). 
Close inspection shows that this is not the case for the Stark spectrum (a). 

A simple possible explanation for the observations could be that the excitation of  Rydberg atoms generally takes place 
at a distance of about 700~$\mu$m from the  trap center where the electrical fields of the Paul trap reach 250~V/cm.
Here, possibly a non-trapped  background gas of Rb atoms could provide enough ion signal. 
In order to test this hypothesis, we displaced the focus of the UV laser so that the laser did not overlap anymore with the cold atom cloud. 
The laser is then  still  able to excite Rydberg states at a distance of 700~$\mu$m from the trap center  but not anymore within the atomic cloud. In our experiments, however,  we observe that the Rydberg signal vanishes. This shows, contrary to the assumption, that most  Rydberg excitations occur within the atomic cloud. 

\begin{figure}
	\includegraphics[width=8cm]{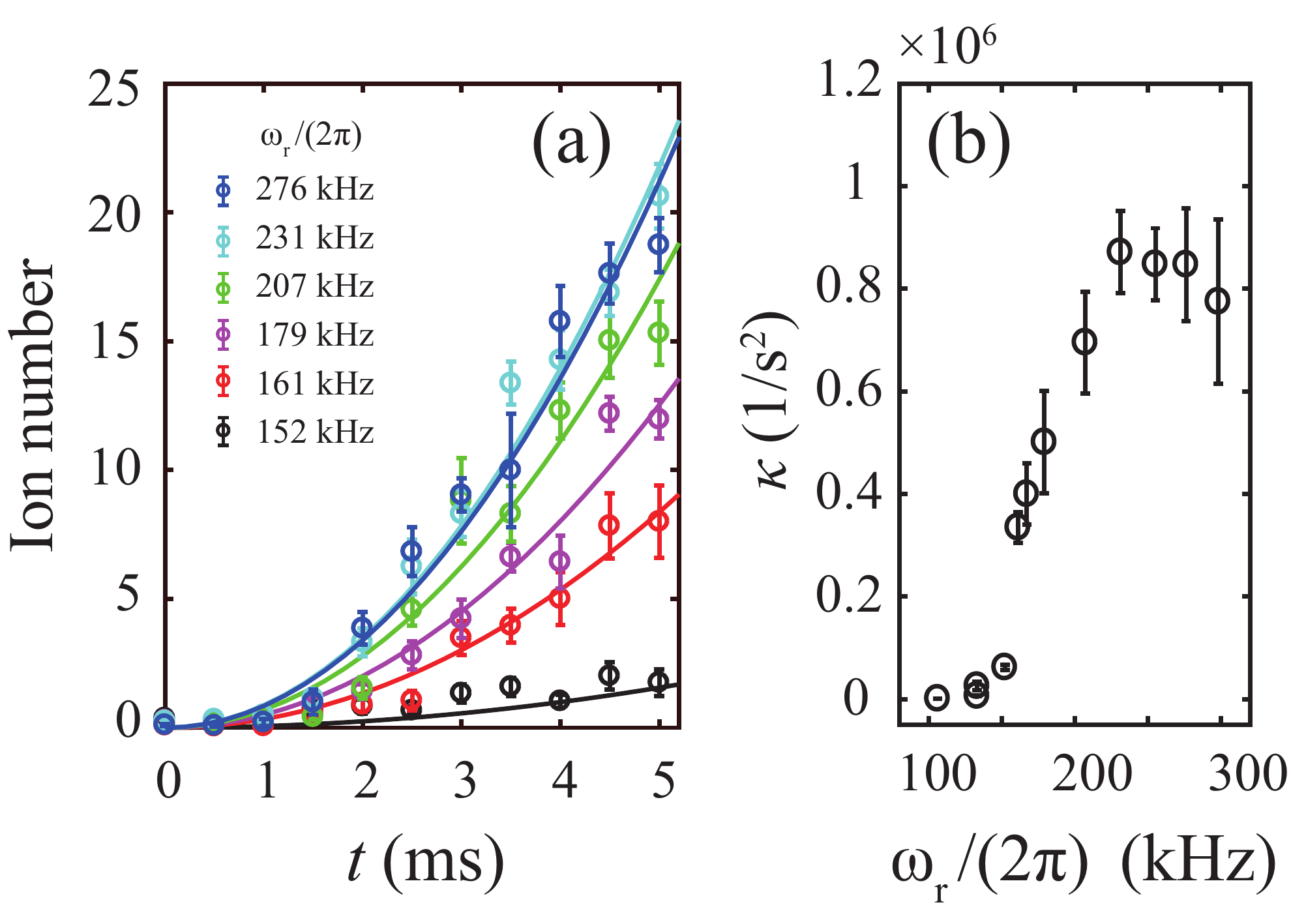}
	\caption{\label{dynamics} (a) Non-linear increase of the number of  trapped ions with time. 
		 Different colors indicate different  radial trapping frequencies $\omega_r$ of the ion trap as given by the legend. The error bars correspond to the statistical uncertainties in the underlying atom number measurements. The solid lines are fits of a quadratic function $\kappa t^2$ to the data. (b) Fit parameter $\kappa$ as a function of $\omega_r$.  The error bars represent the uncertainties as determined by the fits.
	}
\end{figure}

In a second set of experiments, in order to shed more light on the process of Rydberg excitation, we studied the dynamics of the ion production. 
For this, we measured how the number of trapped ions increases as a function of exposure time $t$ to the UV laser.
  Figure \ref{dynamics}(a) shows the results obtained for a fixed laser frequency corresponding to $-69\:\text{GHz}$ and for different radial trapping frequencies of the Paul trap.
   We find a nonlinear increase of the ion number which indicates that the presence of an ion facilitates the  production of  a second one.
   Therefore, this again supports  the notion of a Rydberg atom being excited in the electrical field of an ion. 
   Empirically we find that the data sets are described  well  with the quadratic fit function $\kappa t^2$ (see solid lines), where  $\kappa$ is the fit parameter. 

We find that $\kappa$ generally increases with the trapping frequency, see also Fig.$\:$\ref{dynamics}(b).
This can be explained as follows. By increasing the radial trapping frequency of the Paul trap we increase the ion density  within the atom cloud. 
Again, this goes along with the notion that the ion assists the Rydberg excitation. 
 At radial trapping frequencies larger than $\omega_r\approx 2\pi \times 230\:\text{kHz}$ the parameter $\kappa$ saturates. Possibly there is a reduced compressibility of the ionic cloud due to the Coulomb repulsion. 

In conclusion, we have carried out Rydberg spectroscopy of a mixed atom-ion ensemble within a combined trap. 
By comparing our measured spectra to calculated Stark maps we can infer that large electrical fields of about 250 V/cm
are present. However, we have obtained apparently contradictory results for the  source of these fields. Several measurements indicate
that these fields should originate from ions, however,  the structure of the measured spectrum rather points  towards ion trap fields. 
To solve this question further  investigation is needed.
One possibility is to use a digital  ion trap that  exhibits periods of zero electrical field for the measurements \cite{Ding2002,Bandelow2013,Deb2015}. The spectroscopy is then performed  in a pulsed way at these periods.
This ensures that only electrical fields of ions are available for Stark shifts.
Another possibility is to study the non-linearity coefficient $\kappa$ as a function of laser frequency in order to check  whether its behavior follows the Stark spectrum of a point charge.

We thank Herwig Ott and Florian Meinert for valuable discussions. This work was supported by the German Research Foundation (DFG) within the priority program "Giant Interactions in Rydberg Systems" (DFG SPP 1929 GiRyd). M.D. acknowledges support from Universit\"{a}t Ulm and Ulmer Universit\"{a}tsgesellschaft (UUG) through a Forschungsbonus grant.

\section*{SUPPLEMENTARY INFORMATION}

\subsection*{Effect of the Paul trap RF electrical field}
\makeatletter
\def\p@subsection{}
\makeatother
\setcounter{figure}{0}
\renewcommand{\thefigure}{S\arabic{figure}}

\begin{figure}
\includegraphics[width=8cm]{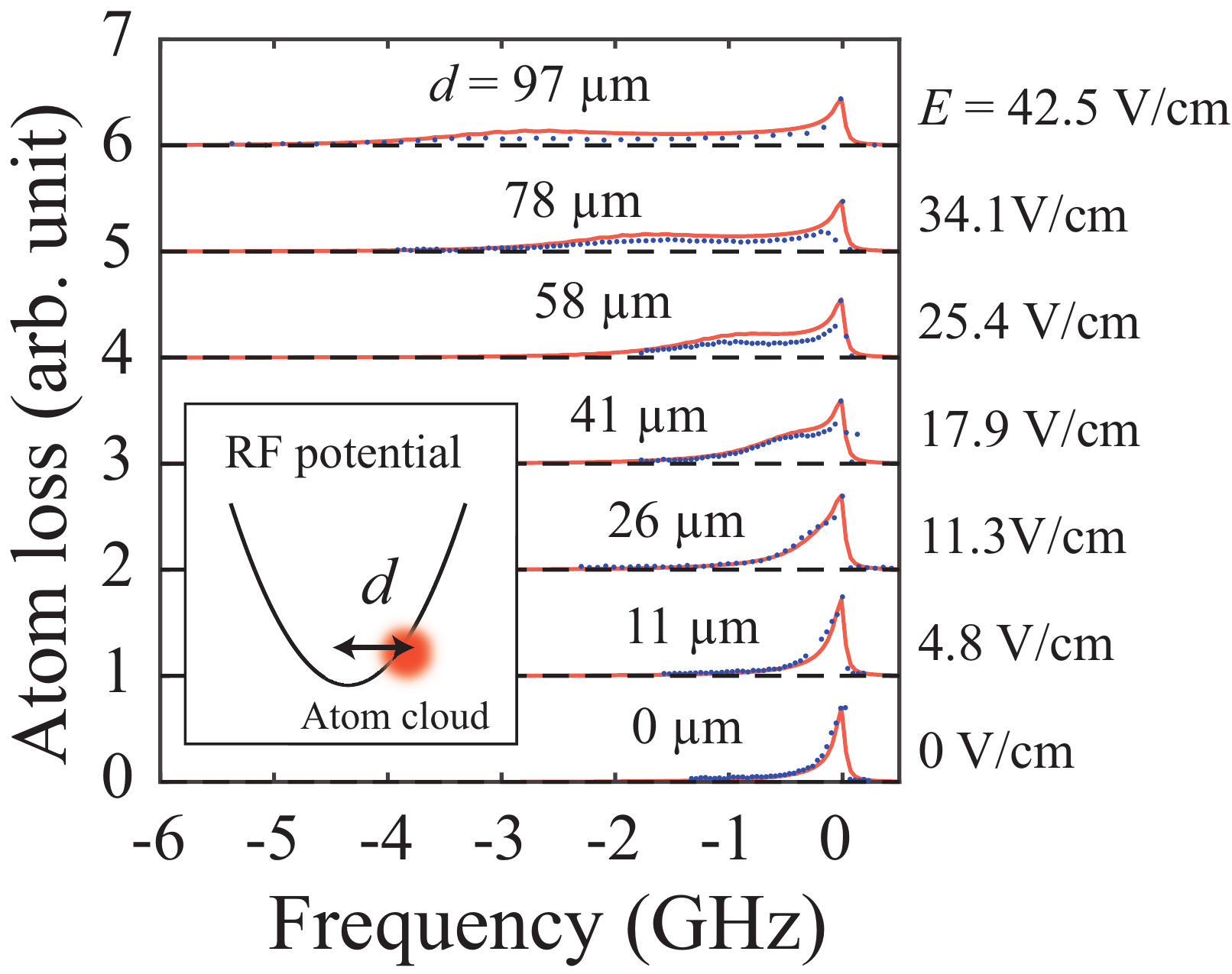}
\caption{\label{RF_broadening} Rydberg spectroscopy in the electrical fields of a Paul trap in the absence of ions. 
	  Shown is the measured atom loss as a function of the frequency of the UV laser light in the vicinity of the  $27P_{3/2}$ resonance line.
	  Spectra are taken  for various distances $d$ between the atom cloud center and the Paul trap center (see inset and legend). The electrical field strength on the right hand side is
	  the RF-field amplitude of the Paul trap at distance $d$. For better visibility the spectra are shifted with respect to each other in vertical direction.
The red solid lines are the results of numerical calculations.
}
\end{figure}

In order to test and verify our general understanding of  Rydberg excitation of atoms in the electric fields of a Paul  trap we have
 carried out the following Rydberg spectroscopy measurement.
 We operate the linear Paul trap with the end cap electrodes turned off
 such that there is no axial confinement, and ions quickly escape from the trap. Thus, we can exclude any atom-ion interactions in the Rydberg spectroscopy. 
  The electric field distribution of the trap is approximately the same as for the trap with the end caps turned on.   
We perform Rydberg spectroscopy in the vicinity of the  $27P_{3/2}$ resonance line.
Here, only atoms are lost that have been excited to the Rydberg state.
Figure \ref{RF_broadening} shows the observed spectra (blue dots).
Each spectrum corresponds to a controlled displacement of the ODT together with the atom cloud from the Paul trap center by a distance $d$, as sketched by the inset in Fig.$\:$\ref{RF_broadening}. Thus, the corresponding atomic clouds experience different electric field amplitudes in their center as 
indicated  on the right hand side of the figure. 
The oscillating and inhomogeneous 
 electrical fields of the Paul trap lead to a broadening and distortion of the  $27P_{3/2}$ resonance line.   
For the maximum value of $d=97\:\mu\text{m}$ presented here, the resonance line spans a frequency interval of about  $4\:\text{GHz}$. 
The red solid lines are the results of  numerical simulations taking into account the temporal and spatial distribution of the electric fields, the spatial distribution of the atoms and the quadratic dependence of the Stark shift on the electrical field. We find good agreement between the measurements and the calculations. This gives us confidence that we have a good general understanding of our experimental conditions.



\end{document}